\documentclass[reprint,aps,prl,twocolumn,superscriptaddress]{revtex4-1}
\usepackage[utf8]{inputenc}
\usepackage{ifpdf}

\usepackage{graphicx}
\usepackage{braket}
\newcommand{\bk}{{\mathbf k}}
\newcommand{\bs}{{\mathbf \sigma}}
\usepackage{amsmath,mathtools}

\begin{document}
\title{Gate-Tunable Quantum Anomalous Hall Effects in MnBi$_2$Te$_4$ Thin Films}
\author{Chao Lei and A.H. MacDonald}
\affiliation{Department of Physics, The University of Texas at Austin, Austin, Texas 78712, USA}

\begin{abstract}
The quantum anomalous Hall (QAH) effect has recently been realized in thin films of 
intrinsic magnetic topological insulators (IMTIs) like MnBi$_2$Te$_4$. Here we point out that that
the QAH gaps of these IMTIs can be optimized, and that both axion insulator/semimetal 
and Chern insulator/semimetal transitions can be driven by electrical gate fields on the $\sim 10$ meV/nm scale.  
This effect is described by combining a simplified coupled-Dirac-cone model 
of multilayer thin films with Schr{\"o}dinger-Poisson self-consistent-field equations.
\end{abstract}

\date{\today}

\maketitle









\textit{Introduction---}
Following its initial experimental realization in magnetically-doped topological insulators (MTI)\cite{Chang2013},
the quantum anomalous Hall (QAH) effect\cite{Haldane1988} has been widely studied\cite{Liu2016,Tokua2019_MTI_review,Bestwick2015,Kou2014_QAH,Kou2015_QAH}.
The QAH effect is of interest because of its potential applications in quantum metrology\cite{Okazaki2020,Gotz2018,Fox2018}
and spintronics\cite{Wu2014,Chang2015}, and because of its potential role as a platform for chiral topological superconductivity\cite{Fu2008,Wang2015}, 
Majorana edge modes\cite{He_2017}, and Majorana zero modes\cite{Zeng_2018}.
Because of strong disorder, thought to be due mainly to random magnetic dopants, 
the QAH effect appears only at extremely low temperatures
in MTIs.  Overcoming this disorder effect has been recognized as key to realizing
the higher temperature QAH effects that would bring more applications within reach.

Topological materials with spatially ordered magnetic moments can be realized by 
forming heterojunctions between ferromagnetic insulators and topological 
insulators\cite{Tokua2019_MTI_review,Wei2013,Lee2016,Katmis2016,Lang2014_YIG} or by growing 
intrinsic MnBi$_2$X$_4$ or MnSb$_2$X$_4$ magnetic topological insualtors (IMTIs) and related superlattices \cite{Lei2020,Otrokov_2017,Eremeev2017,Otrokov2019,Zhang2019,Li2019_theory,Chowdhury_2019,Lee2013,Rienks2019,Zeugner2019,Yan2019,Lee2019,Li2020,Otrokov2019_film,Liu2020,Chen2019_Pressure,Deng2020,Deng_2020,Gong2019,Zhang2019_AHC,Li2019,Hao2019,Chen2019,Ge2020,Hu2020,Ding2020,Lee2019,Swatek2020,Eremeev2018,Wu2019,Vidal2019,Klimovskikh2019,Sun2019,Gu_2020,Wimmer2020,Belopolski2017}, where X=(Se,Te).
The IMTIs consist of van der Waals coupled septuple-layer building blocks that have Mn local moment layers at their centers.
To date the anomalous Hall resistances measured\cite{Tokua2019_MTI_review,Wei2013,Lee2016,Katmis2016,Lang2014_YIG}
in the heterojunction systems are still far from their ideal quantized values, 
mainly due to weak exchange coupling between the surface states of the topological insulator and moments in the ferromagnetic 
insulator.   On the other hand reasonably accurately quantized Hall resistances have been measured\cite{Deng2020} in
five-septuple layer thin films of MnBi$_2$Te$_4$ (MBT) in the absence of magnetic field
at a temperatures exceeding $1$K and, in the presence of magnetic field $\sim 5$T, at
other film thicknesses\cite{Deng2020,Ge2020,Liu2020} and higher temperatures.
Although the QAH temperature is larger in the IMTI case than in the magnetic-dopant 
case, the anomalous Hall resistances $R_{xy}$ can deviate by as much as a factor of 3$\%$ from exact quantization
and the longitudinal resistances $R_{xx}$ are still $\sim 0.01-0.02 h/e^2$.
This compares with Hall resistivity deviations smaller than 1 ppm \cite{Gotz2018,Fox2018} at the lowest temperatures in 
the magnetic-dopant case.  In this paper we theoretically explore the possibility of optimizing the QAH in MBT by 
applying electrical gate voltages to increase the QAH gap, and also address gate-tuned transitions between
insulating and semimetallic states, and in the case of ferromagnetic spin configurations 
between insulating states with different Chern numbers. 

\ifpdf
\begin{figure}[htp]
\centering
\includegraphics[width=0.9\linewidth,height=0.65\linewidth]{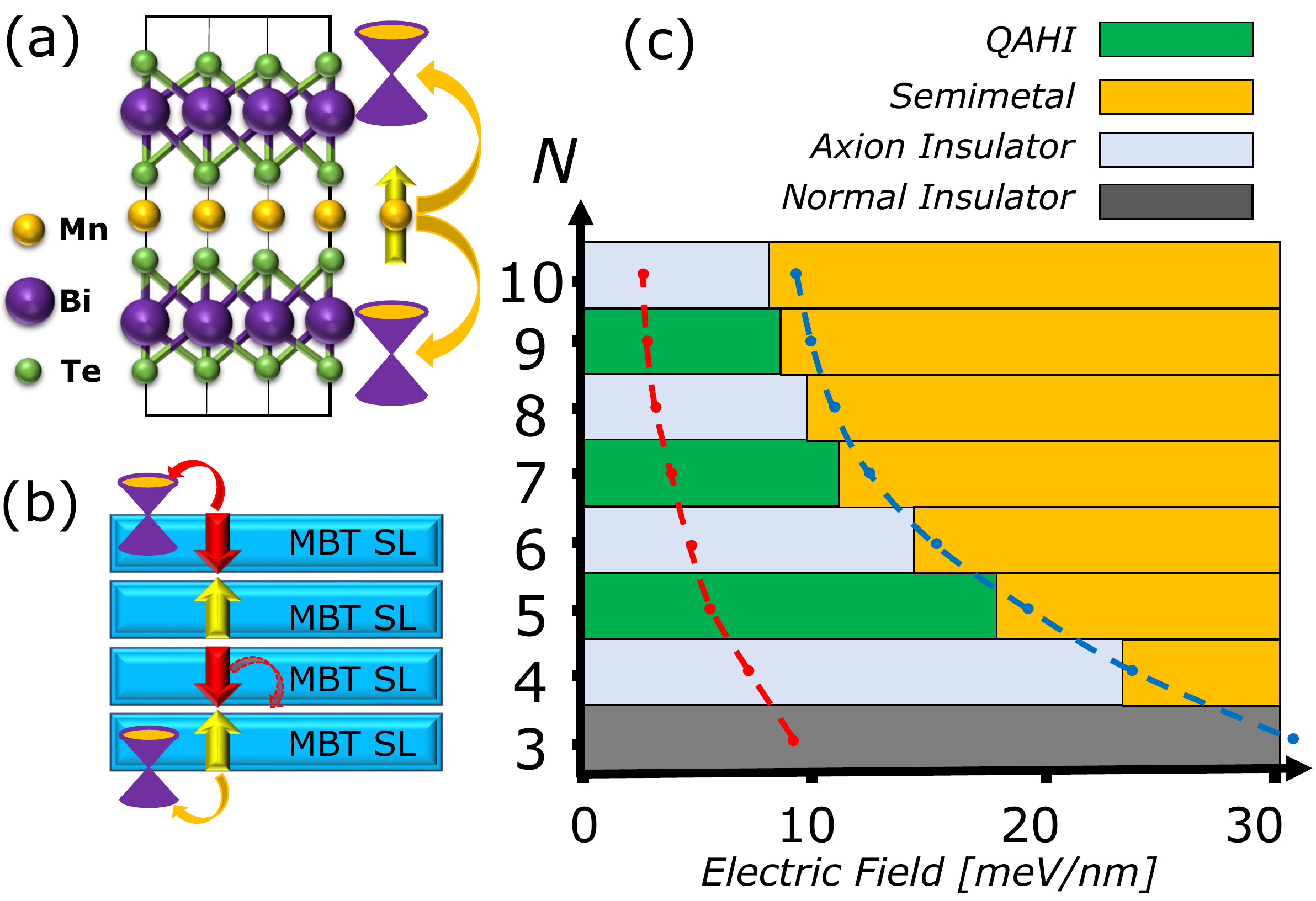}
\caption{
(a) Crystal structure of one septuple layer of MBT, which consists of seven layers with one magnetic Mn ion
in the center and two Te-Bi-Te trilayers outside; Two Dirac cones lie at the surface.
(b) Illustration of the coupling between the Dirac cones at the two surfaces of 
antiferromagnetic MBT thin films.  The Dirac-cone masses are produced by exchange coupling to 
Mn local moments and are opposite in sign for even layer-number thin films and identical for odd-number-layer thin films.
(c) Phase diagram of antiferromagnetic thin films, with electric field on the $x$ axis and the number of septuple layers N
in the film along the  $y$ axis. Different colors represent different phases. 
The red and blue dashed lines respectively plot the electric fields $E$ at which $eEt_N=E_{gs}$ and 
$eEt_N/\epsilon_{zz}=E_{gs}$, where $E_{gs} \sim 37$ meV is the surface state gap at $E=0$ \cite{Lei2020} and $\epsilon_{zz} \sim 3.5$ 
is the perpendicular component of the static dielectric constant of the coupled Dirac cone model.(See main text.)
} \label{fig:phase}
\end{figure}
\fi

Our analysis is based on the simplified coupled Dirac-cone model\cite{Lei2020}
illustrated schematically in Fig. \ref{fig:phase} (a) and (b)) that captures most topological and electronic 
properties, and on a self-consistent-field Sch{\"o}dinger-Possion approximation for the interacting carriers.
We show that gates can maximize the QAH gap either by compensating for unintentional electric fields, or
in the case of high-Chern-number ferromagnetic (FM) insulators, by tuning the gate field to an optimal non-zero value.

\textit{Gate-Field Phase Diagram---} The low-energy properties of MBT thin films are 
accurately modeled \cite{Lei2020,Burkov_Balents_2011} by a simple Hamiltonian 
that includes only Dirac cone surface states on the top and bottom surface of each septuple layer (as illustrated in Fig. \ref{fig:phase} (a)), 
and hopping between Dirac cones:
\begin{widetext}
\begin{equation}
\label{eq:model}
   H =  \sum_{\bk_{\perp},ij} \Big[\Big( \, 
   (-)^i  \hbar v_{_D}  (\hat{z} \times \bs) \cdot \bk_{\perp} + m_{i} \sigma_z + V_i \Big) \delta_{ij}  
    + \Delta_{ij}(1-\delta_{ij} ) \Big] c_{\bk_{\perp} i}^{\dagger} c_{\bk_{\perp} j} ~.
\end{equation}
\end{widetext}
Here the Dirac cone labels $i$ and $j$ are respectively odd and even on the top and 
bottom surface of each septuple layer, $\hbar$ is the reduced Planck's constant, 
$v_{_D}$ is the Dirac-cone velocity and $V_i$ is the self-consistent Hartree potential on surface $i$.
In the following we retain only the largest Dirac-cone hybridization parameters, 
letting $\Delta_{ij} \to \Delta_{S}$ for hopping within the same septuple layer
and $\Delta_{ij} \to \Delta_{D}$ for hopping across the van der Waals gap between adjacent septuple layers.
Exchange interactions between Dirac-cone spins and local moments in the interior of each 
septuple layer are captured by the mass gap parameter
$m_{i} = \sum_{\alpha} J_{i\alpha} M_{\alpha}$ where $ \alpha $ is a septuple-layer label and  
$M_{\alpha} = \pm 1$ specifies the sense of magnetization on layer 
$\alpha$. We include interactions between Dirac cone spins and Mn local moments in the same septuple layer
with exchange constant $J_S$ and with local moments in the closest adjacent layer with 
exchange constant $J_D$.  In this paper we set $\Delta_S$=84 meV, $\Delta_D$ = -127 meV, $J_S$ = 36 meV and $J_D$ = 29 meV,
based on the fit to MnBi$_2$Te$_4$ {\it ab initio} electronic structure calculations discussed in 
detail in Ref.~\onlinecite{Lei2020}.

The Dirac-cone Hartree potentials $V_i$ in Eq.~\ref{eq:model} are calculated from a discrete Poisson equation
in which positions $z_i$ are assigned to Dirac-cone states ordered sequentially from top to bottom.
The position assignments are based on microscopic charge-density-weighted average positions
discussed in the Supplementary Material\cite{SI}.  The discrete Poisson equation reads 
\begin{equation}
\begin{split}
    & \tilde{\epsilon} \mathcal{E}_i = \tilde{\epsilon} \mathcal{E}_{t} + 
    \sum_{j=1}^{i} \delta \rho_i  = \tilde{\epsilon} \mathcal{E}_{i-1} + \delta \rho_i \\
    & V_i = \sum_{j=2}^{i} \mathcal{E}_i (z_i-z_{i-1}) = V_{i-1} + \mathcal{E}_i  (z_i-z_{i-1}) \,\, .
\end{split}
\end{equation}
Here $\mathcal{E}_t=\mathcal{E}_0$ is the electric field controlled by the top gate
above the top surface of the thin film, $\mathcal{E}_i$ is the electric field between surface $i$ and $i+1$,
$\tilde{\epsilon}$ is intended to account for gate field 
screening by degrees of freedom not included in our model,
and $V_i$ and $\delta \rho_i$ are the Hartree-potential and net surface charge density at surface $i$.  (We choose $V_1=0$.)
The bulk perpendicular dielectric constant $\epsilon_{zz}$ of MBT has not been measured, to our knowledge, but should be close 
to $\epsilon_{zz} \sim 3$ measured in Bi$_2$Te$_3$ \cite{Dheepa2005_dielectric}.
By calculating the imaginary part of the conductivity of the bulk limit of the Dirac cone 
model \cite{SI}, we find that its bulk dielectric constant $\epsilon_{zz} \sim 3.5$.
We have therefore concluded that most of the perpendicular screening in MBT is captured by the 
Dirac cone model and set $\tilde{\epsilon}=1$ in all the explicit calculations we describe. 
The surface charge densities used in the Poisson equation are calculated self-consistently from the 
electronic structure model using 
\begin{equation}
    \rho_i = -e \int \frac{d\bk}{(2\pi)^2} \sum_{n=1}^{4N} \sum_{s = \uparrow, \downarrow} |\Psi_{n\bk}^{\sigma}(z_i)|^2 f(E_{n\bk}-\mu), 
    \label{eq:chargedensity}
\end{equation}
where $e$ is electron charge, $n$ is a quasi-2D band label, $f(E_{n\bk}-\mu)$ is the Fermi-Dirac distribution function, and $\mu$ is the chemical potential.  The net surface charge density $\delta \rho_i$, which appears in the Poisson equation, 
is defined as the difference between $\rho_i$ calculated from Eq.~\ref{eq:chargedensity} and the charge density calculated 
from the same equation with the Fermi level in the gap and no gate field.  This prescription is motivated by the 
linearity of the Poisson equation, and by the fact that the bare bands have been fit to the electronic structure of 
neutral ungated thin films.

For a thin film with $N$ septuple layers, the model has $2N$ Dirac cones located at the surfaces
of each septuple layer and $4N$ bands, $2N$ of which are occupied 
at neutrality and temperature $T=0$ -- the limit considered in this paper.
The electric field below the bottom layer of the 
thin film $\mathcal{E}_b = \mathcal{E}_t $ at neutrality. The antiferromagnetic (AF) thin film phase diagrams obtained
from these self-consistent Hartree calculations are summarized in Fig. \ref{fig:phase} (c), from which it follows
that the thin films become semimetals when the electric field exceeds a critical value. 
To test the reliability of the simplified electronic structure model in accounting for gate-field 
response, we performed corresponding DFT calculations of the electron structure of MBT thin film with 2-4 septuple layers.
The critical fields using the two approaches are in qualitative agreement.\cite{SI}.
The critical electric field at which the semimetallic state is reached
decreases when the thickness of thin film increases and asymptotically approaches 
$E_{gs}\epsilon_{zz}/t_N$, where $E_{gs} \sim 37$ meV is the surface state energy gap \cite{Lei2020}
and $t_N$ the thickness of an $N$ septuple layer film.
At small electric fields, the thin films exhibit an even-odd effect, 
namely that thin films with an even number of septuple layers are axion insulators with 
strong magneto-electric response properties~\cite{Zhang2019,Zhu2020},
whereas those with an odd layer-number ($N>3$) are QAH insulators \cite{Lei2020}. 
This odd-even effect is illustrated schematically in Fig. \ref{fig:phase} (b), where it can be seen that the
Dirac cones at the top and bottom surfaces have opposite masses for even layer-number, but have 
identical masses for odd layer-number, this characteristic can be modeled with a effective toy model \cite{SI}.

\textit{Gate-control of the QAH effect ---} 
In antiferromagnetic MBT thin films quantized anomalous Hall resistances are 
expected when the number of septuple layers $N$ is odd.  Under these circumstances 
the film has residual magnetism due to uncompensated moments,
and the film is sufficiently thick \cite{Lei2020} that coupling between top and bottom surfaces does not induce a 
transition to a trivial insulator.  DFT calculations\cite{Otrokov2019_film} predict that the 
QAH effect occurs for three or more layers, and a robust 
QAH effect has been measured in high-quality five-septuple-layer MBT 
thin films\cite{Deng2020}.  

\ifpdf
\begin{figure}[htp]
\centering
\includegraphics[width=0.9\linewidth,height=0.9\linewidth]{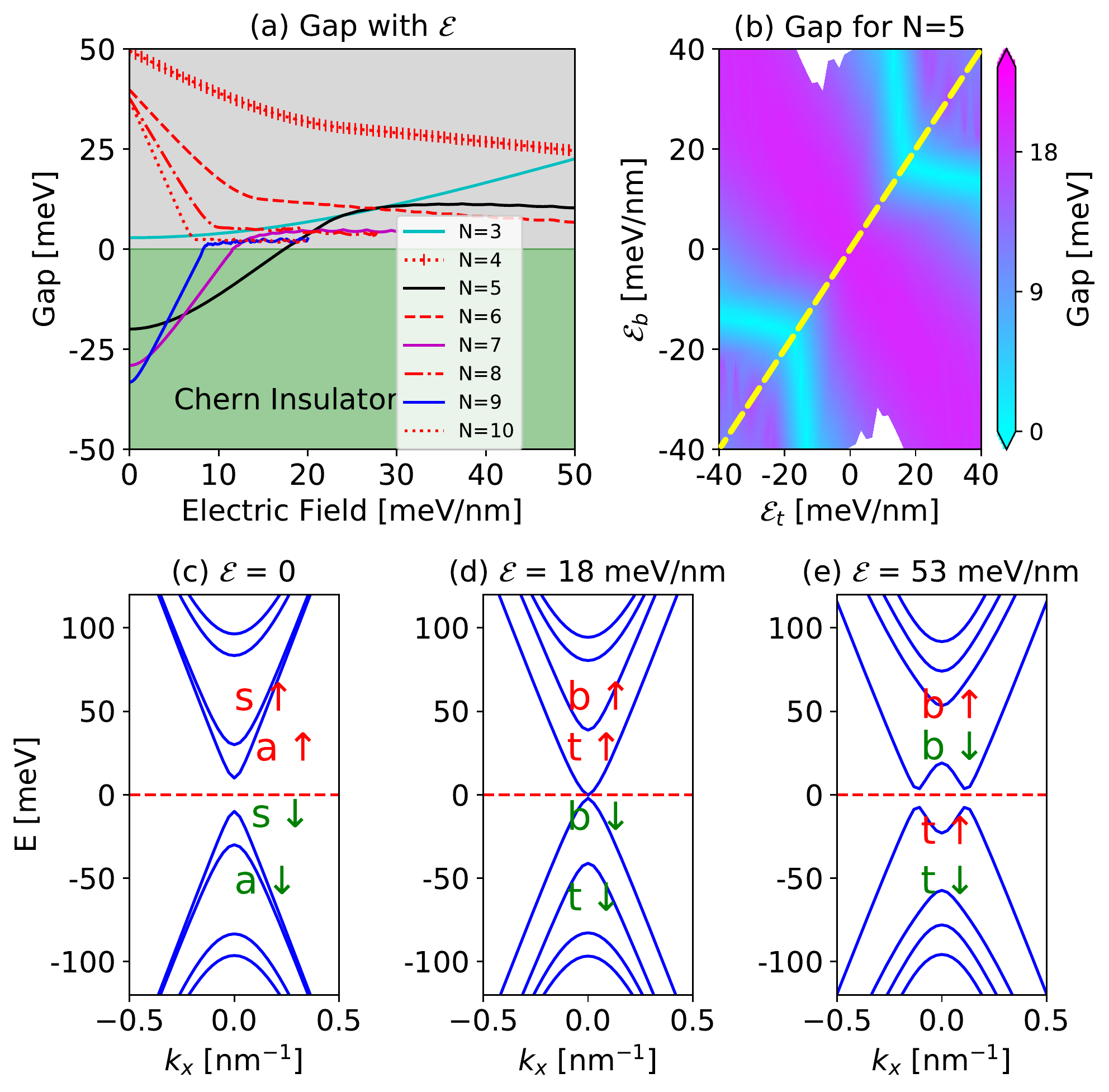}
\caption{
(a) Gap {\it vs.} gate electric field for $N$-septuple layer charge-neutral antiferromagnetic thin films. A negative sign is assigned to the gaps when the Chern number is non-zero.  The green shaded regions contain Chern insulator phases while the 
grey shaded regions contain trivial insulator or semimetal phases (see text).
(b) Gap {\it vs.} gate electric fields for five-septuple-layer thin films.  
Neutrality occurs along the yellow dashed line.
(c)-(d) Self-consistent bandstructures of neutral five-septuple-layer thin films.  
In the absence of an electric field bands are labelled as layer symmetric (s) or 
anti-symmetric(a) and by majority spin ($\uparrow$ or $\downarrow$).  
Bands at finite electric field are labelled by majority surface ($t$ for top or $b$ for bottom) and majority spin.}
\label{fig:bands}
\end{figure}
\fi

In Fig.\ref{fig:bands} (a) we plot the charge gap of several MBT thin films
{\it vs.} gate electric field.  For even $N$ all insulators are trival, and gaps decrease with gate fields.
The parameters we have chosen for the coupled Dirac-cone model
place the N=3 thin film on the trivial side of the topological phase transition,  
but an increase of the magnetic exchange coupling parameters by as little as  
several meV would drive the system from a trivial insulator state to a Chern insulator state. 
If the ideal ungated three-setptuple layer antiferromagnetic MBT thin film is indeed a trivial insulator,
our calculations show that a gate field would not be able to drive the system into a Chern insulator state
since the condition to be a Chern insulator is $m> \sqrt{\Delta^2 + V^2}$ \cite{SI} in the presence of electric field.
Here $m$,$\Delta$,$V$ are the parameters the of mass, hybridization of top and bottom surface states, 
and Hartree potential induce by the electric field in the toy model illustrated in the supplemental material.
On the other hand a small electric field due to asymmetric unintended doping would close the gap of   
the Chern insulator state~\cite{SI}, even if it were stable in the ideal case.
In contrast odd $N$ films with $N \geq 5$ all have robust QAH effects in the absence of a gate field
as shown in Fig. \ref{fig:bands} (a) where we see that gate fields always act to reduce thin film gaps.
The ideal QAHE gap cannot be enhanced by gate fields, but dual gating will still be valuable in 
practice since real samples normally have unintended electric fields for which the gate fields can compensate.  
This behavior can be understood qualitatively as a consequence of an energetic shift of the magnetically
gaped states on one surface relative to those on the other surface, so that the conduction band states of the
low-potential energy surface falls below the top of the valence band on the high-potential energy
surface. The gaps tend to survive to larger gate electric fields for
even $N$ than for odd $N$ because the approaching valence and conduction band extrema states have the same 
dominant spin in the former case, strengthening level repulsion effects.  In the limit of thick films the critical electric field 
$E$ required to close the gap approaches the value $E_{gs}\epsilon_{zz}/et_N$ where $E_{gs} \sim 37$ meV is the energy gap
of isolated surface states.  

The energy gap in the quasi-2D band structure of the $N=5$ thin film
is plotted {\it vs.} top and bottom gate fields
in Fig.~\ref{fig:bands} (b), where the yellow dashed line marks the neutrality line.
We see here that as the carrier densities of the films thicken, the overall gap is 
controlled more and more by independent screening of external electric fields by carriers
near either surface.  When carriers are present, the screened electric field drops toward 
near the middle of thick films and larger electric fields are generally required to close the gap.
Of course, when the Fermi level does not lie in the gap and away from the neutrality line, the 
resulting states are magnetically ordered two-dimensional Fermi liquids with large 
momentum-space Berry curvatures \cite{Xiao2010}, not Chern insulators.
Because these itinerant electron ferromagnets are strongly gate tunable, they are potentially interesting 
for spintronics.  

The band structure evolution with gate field for $N=5$ neutral antiferromagnets is 
illustrated in Fig. \ref{fig:bands} (c)-(e).  In these plots we have labeled the subbands based on the 
projection of their densities-of-states to 
individual spins, and to Dirac cones associated with particular septuple layers.
For $N=5$ antiferromagnets the Hamiltonian in the absence of a gate field (Fig. \ref{fig:bands} (c)) 
possesses a $z \to -z$ mirror symmmetry which leads to band 
eigenstates that are either symmetric ($s$) or antisymmetric ($a$) under this symmetry operation. 
At finite gate fields the four subbands around the Fermi level reside mainly on the top or bottom 
surfaces ($t$ for top and $b$ for bottom) of the thin film 
and that they are strongly spin-polarized ($\uparrow$ or $\downarrow$). 
Either $t\uparrow$ and $b\downarrow$ or $t\downarrow$ and $b\uparrow$ subbands
lie close to the Fermi level depending on the direction of the electric field and the 
spin-configuration. Fig. \ref{fig:bands} (d) shows the bands near the critical value 
at which band touching first occurs, near $18$ meV/nm for $\epsilon_{zz} \sim 3.5$.
At larger fields (Fig. \ref{fig:bands} (e)), the $b\downarrow$ and $t\uparrow$ subbands are inverted. 
The inversion changes the polarizations of the subband just below Fermi level, 
and drives the thin film from a state with Chern number $C=1$ to semimetal state with $C=0$.
In the semimetal state beyond the critical electric field, 
a small gap reopens in our simplified electron-structure model due to weak-coupling between
top and bottom surfaces.  These small gaps are not expected to survive the anisotropic band 
dispersion of more fully realistic models.

\textit{High-Chern-number QAH Systems---}
Because their antiferromagnetic interlayer exchange interactions are exceptionally weak,
the Mn local moment spins in MBT are aligned by external magnetic fields larger than $\sim 5$ Tesla \cite{Deng2020,Ge2020,Liu2020}.  Since bulk ferromagnetic MBT is a Weyl semimetal, in the thick film limit, 
the Chern number per layer of a ferromagnetic MBT film approaches the dimensionless Hall conductivity per layer 
of the bulk semimetal ($\sim 0.2$), with finite-size energy gaps that tend to get smaller as the films get thicker. 
The Chern number of ferromagnetic MBT thin films with $N > 8$ 
are larger than 1 \cite{Ge2020,Lei2020} in the absence of a gate field.  
In Fig. \ref{fig:chern} (a) we show the gaps {\it vs.} gate electric fields for FM thin films 
with from with $N$ from 7 to 10, crossing the thickness at which the Chern number jumps from 1 to 2 
at zero electric field.  Overall the gaps tend to decrease with gate field as in the antiferromagnetic case.
An exception occurs for the $N=9$ and $N=10$ films, for which the gaps initially increase as the 
system moves further from the $C=2$ to $C=1$ boundary with gate field.  In Fig. \ref{fig:chern} (b), where we assign a positive sign to the gap of  odd-Chern-number(C=1) states and a negative sign for even-Chern-number(C=0 or 2) states, we illustrate the thickness 
dependence for a series of gate fields.  Gate electric fields can induce 
transitions between insulators with different Chern numbers as illustrated in 
Fig.~\ref{fig:chern} (c). 

\ifpdf
\begin{figure}[htp]
\centering
\includegraphics[width=0.9\linewidth,height=0.9\linewidth]{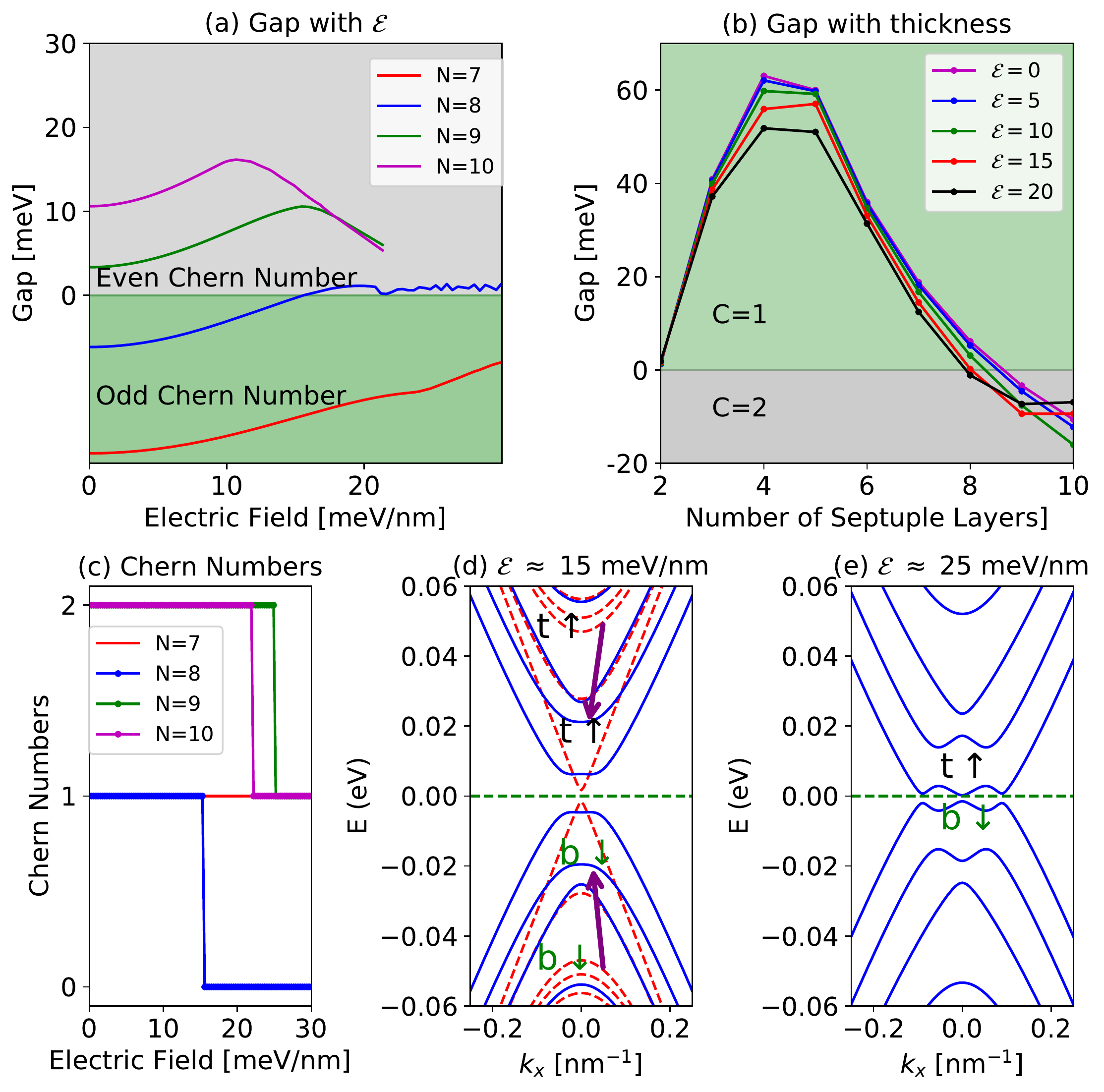}
\caption{
(a) Gaps {\it vs.} gate electric fields for ferromagnetic thin films from 7 to 10 septuple layers. 
For the $N=7$ thin film no gap closing happens occurs below 30 meV/nm, whereas 
for the $N=8$ thin film, the QAH gap closes at $\mathcal{E} \sim 17$ meV/nm, 
For $N=$9/10 thin films, the QAH gap closes at $\mathcal{E} \sim 25$ meV/nm;
(b) QAH gaps {\it vs.} thickness for several electric fields from $\mathcal{E} = 0$ to $\mathcal{E} = 20$ meV/nm;
(c) Chern numbers {\it vs.} electric field for several film thicknesses from $N=7$ to $N=10$; 
(d) and (e) are bandstructure of $N=9$ ferromagnetic thin film in different electric fields.
} \label{fig:chern}
\end{figure}
\fi

The green and magenta curves for $N=9$ and $N=10$ in Fig. \ref{fig:chern} (a) show that the 
maximum QAH gaps are reached at around 10-15 meV/nm.   
At around 25 meV/nm, a topological phase transition occurs at which the Chern numbers change from 2 to 1, and further change to 0 as the electric fields increase (shown as in Fig. \ref{fig:chern} (c)).  These transitions are 
illustrated further in Fig. \ref{fig:chern} (d) and (e), where we show the bandstructures of 
FM thin films with $N = 9$ in  Fig. \ref{fig:chern} (d) and (e), with zero electric field (red dashed curve in (d)), with electric field $\mathcal{E} = 15 $ meV/nm (blue curve in (d)), and with electric field $\mathcal{E} = 25 $ meV/nm (in (e)). 
Unlike the bandstructures of the AF thin film with $N=5$ in Fig.\ref{fig:bands}, the bands closest to the 
Fermi level are not located primarily in the $t\uparrow$ and $b\downarrow$ septuple layers and 
are instead spread across the thin film.  As the electric field increases, the $t\uparrow$ subband is pushed down 
and the $b\downarrow$ subbands is pulled up. Before these two subbands touch up at $\mathcal{E} \sim 25$ meV/nm (shown in Fig. \ref{fig:chern}(e)), hybridization between the electron and hole subbands spread across the entire thin film increases and thus increase the QAH gap. When the $t\uparrow$ and $b\downarrow$ subbands touch at $\mathcal{E} \sim 25$ meV/nm, a topological phase transition occurs
at which the Chern number changes from 2 to 1. Similar band inversions occur for other two subbands which are not shown in the figure, and eventually change the Chern number from 1 to 0.

\textit{Discussion---}
In this paper we have focused on neutral MBT thin films.  Gate-tuning will be 
phenomenologically richer in the case of electrostatic doping, where the ground states 
are expected to be extremely tunable magnetically ordered two-dimensional metals. 
The Hall effect will remain quantized in doped samples, provided that the added 
charges are localized.  Since magnetization textures are charged in Chern insulators,
we anticipate the possibility of engineering Skyrmion lattice ground states\cite{Lee1990,Sondhi1993,Moon1995,Fertig1994} at finite
doping.  Separately, gate fields can be used to engineer strong spin-orbit coupling\cite{You2020},
which is ubiquitous, in Fermi liquid states and to control the interplay between
the itinerant electron and Mn local moment contributions to the magnetization.

In summary, we have studied gate tuning effects in MBT thin films using self-consistent Schr{\"o}dinger-Poisson equations, demonstrating that gates can optimize the QAH gap either by compensating for unintentional electric fields, or in the case of high-Chern-number ferromagnetic (FM) insulators, by tuning the gate field to an optimal non-zero value. Our theory provide an explanation for the absence of the QAH effect in AF thin film with 
three septuple layers, and sheds light on strategies to optimize the QAH effect in both antiferromagnetic and ferromagnetic MBT
thin films using gates. The gate electric fields that induce large changes are 
on the scale of 10 meV/nm,  which is easily realized experimentally.


\section{Acknowledgements} This work Research was sponsored by the Army Research Office under Grant Number W911NF-16-1-0472,and by the Welch Foundation under grant Welch F-1473. We acknowledge helpful discussions with Anton Burkov, Gaurav Chaudhary, Paul Haney, Olle Heinonen, Rob McQueeney, Suyang Xu and Fei Xue. 
The authors acknowledge the Texas Advanced Computing Center (TACC) at The University of Texas at Austin 
for providing HPC resources that have contributed to the research results reported within this paper. 
\bibliography{mbt}

\clearpage
\onecolumngrid
\appendix
\begin{center}
{\bf{Supplementary Material for \\''Gate-Tunable Quantum Anomalous Hall Effects in MnBi$_2$Te$_4$ Thin Films''}}
\end{center}
\setcounter{equation}{0}
\setcounter{figure}{0}
\setcounter{table}{0}
\setcounter{page}{1}
\makeatletter

\newcommand*{\dprime}{\prime\prime}
\newcommand*{\trprime}{\prime\prime\prime}
\renewcommand{\theequation}{S\arabic{equation}}
\renewcommand{\thefigure}{S\arabic{figure}}
\renewcommand{\thesection}{S\arabic{section}}
\renewcommand{\bibnumfmt}[1]{[S#1]}
\renewcommand{\citenumfont}[1]{S#1}

\tableofcontents
\newpage

\begin{appendix}
  \listoffigures
\end{appendix}

\newpage

\section{DFT calculations}
The density functional theory (DFT) calculations were
performed using the Vienna Ab initio simulation package (VASP)~\cite{VASP1993,VASP1994,VASP1996a,VASP1996b}
using Generalized Gradient Approximation PBE~\cite{PBE1996,PBE1997} pseudopotentials.
For the  plane wave expansion of the
DFT calculation we used a cutoff energy of 600 eV, a total electronic energy
convergence threshold of $10^{-7}$ eV per unit cell,
a $9 \times 9 \times 1$ k mesh is used for self-consistent calculations.
A gaussian smearing parameter of 0.05 eV was used to assign partial occupancies.

\section{Discretization of Poisson Equation}
The Dirac-cone Hartree potentials $V_i$ in main text are assigned with positions $z_i$ which are based on microscopic charge-density-weighted average positions:
\begin{equation}
 z_i = \sum_j w_j z_j
\end{equation}
with $j = \rm Bi,Te$ and $w_j = p_j/\sum_j p_j $, where $p_j$ is the projected density of states for Bi and Te atoms besides Mn ion(shown in Fig. \ref{fig:dz} for 3-6 septuple layers MBT) got from density-functional-theory.

\ifpdf
\begin{figure}
\centering
  \includegraphics[width=0.9 \linewidth ]{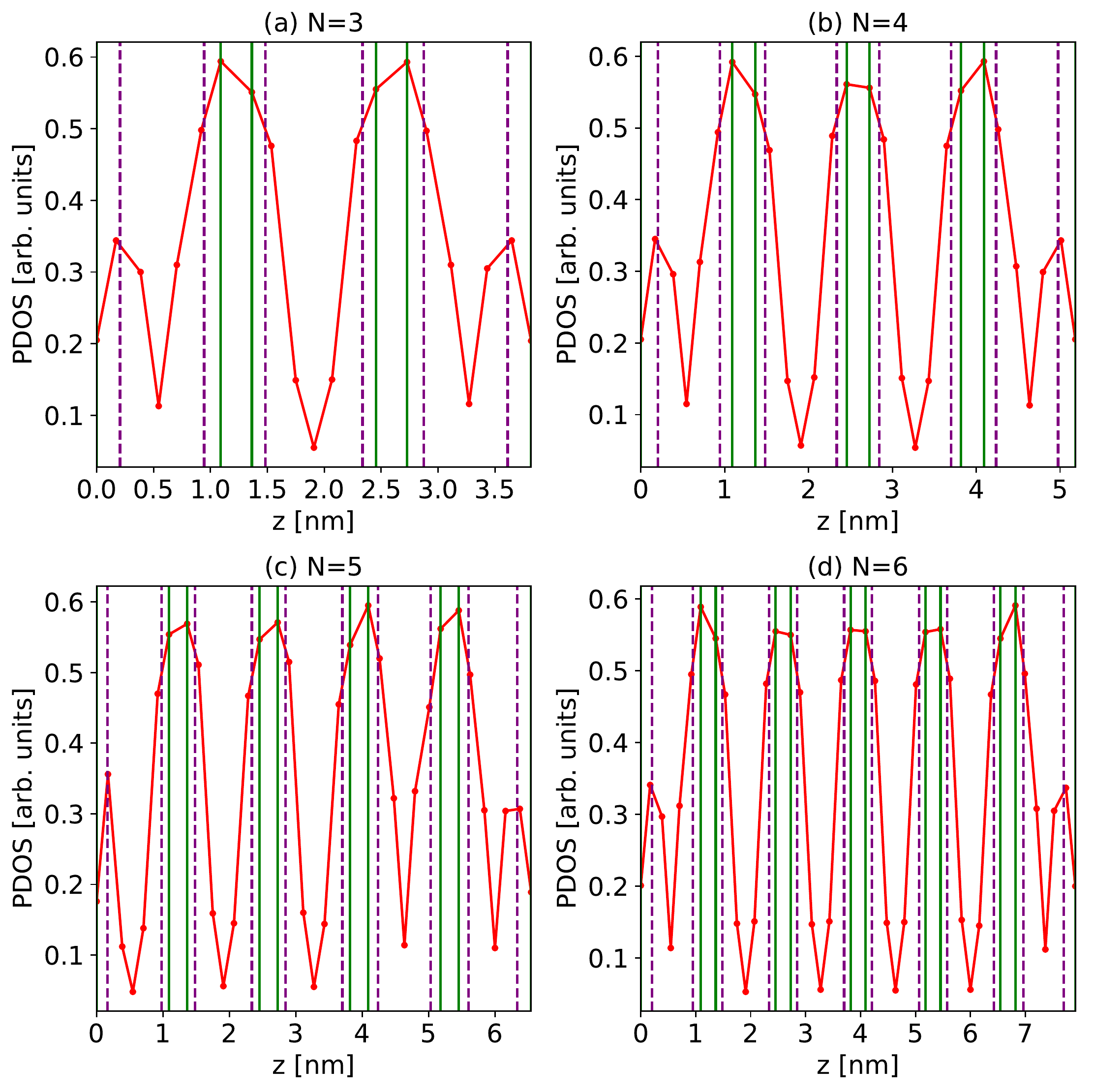}
  \caption{Projected density of states of antiferromagnetic MnBi$_2$Te$_4$ thin films with thickness of 3-6 septuple layers. The green solid vertical lines are labeling the outer Te atoms of each septuple layers, while the magenta dashed lines are the weighted location of each Dirac cones.
  }\label{fig:dz}
\end{figure}
\fi

\vspace{5mm} 
\section{Calculation of bulk dielectric constant}
The bulk dielectric constant is calculated based on the optical conductivity:
\begin{equation}
    \epsilon_{zz} = \lim_{\omega \rightarrow 0} (1 + 4i\pi \frac{\partial \sigma_{zz} }{\partial\omega}),
\end{equation}
with the optical conductivity as:
\begin{widetext}
\begin{equation}
    \sigma_{zz}(\omega) = \frac{ie^2}{\hbar} \int \frac{d\bk}{(2\pi)^3} \sum_{nm} \frac{f_{n\bk} - f_{m\bk}}{E_{n\bk} - E_{m\bk}} \frac{ \braket{\psi_{m\bk} | \partial_z H_{\bk} | \psi_{n\bk} } \braket{\psi_{n\bk} | \partial_z H_{\bk} | \psi_{m\bk} } } {E_{n\bk} - E_{m\bk} -(\hbar \omega + i\eta)},
\end{equation}
\end{widetext}
where $\omega$ is the frequency and $f_{n\bk}$ is the Fermi-Dirac function.

\section{Trilayer MBT thin film}
A topological phase transition happens at $J_S \approx 39~ meV$ for the N=3 thin film in the coupled Dirac-cone model, the parameters we have chosen place the thin film on the trivial side, in Fig. \ref{fig:gap_trilayer} we see that the electric field required to drive the N=3 thin film depends on the value of $J_S$, and a small electric field would destroy the Chern insulator state when the film is a Chern insulator.

\ifpdf
\begin{figure}[htp]
  \centering
  \includegraphics[width=0.9 \linewidth ]{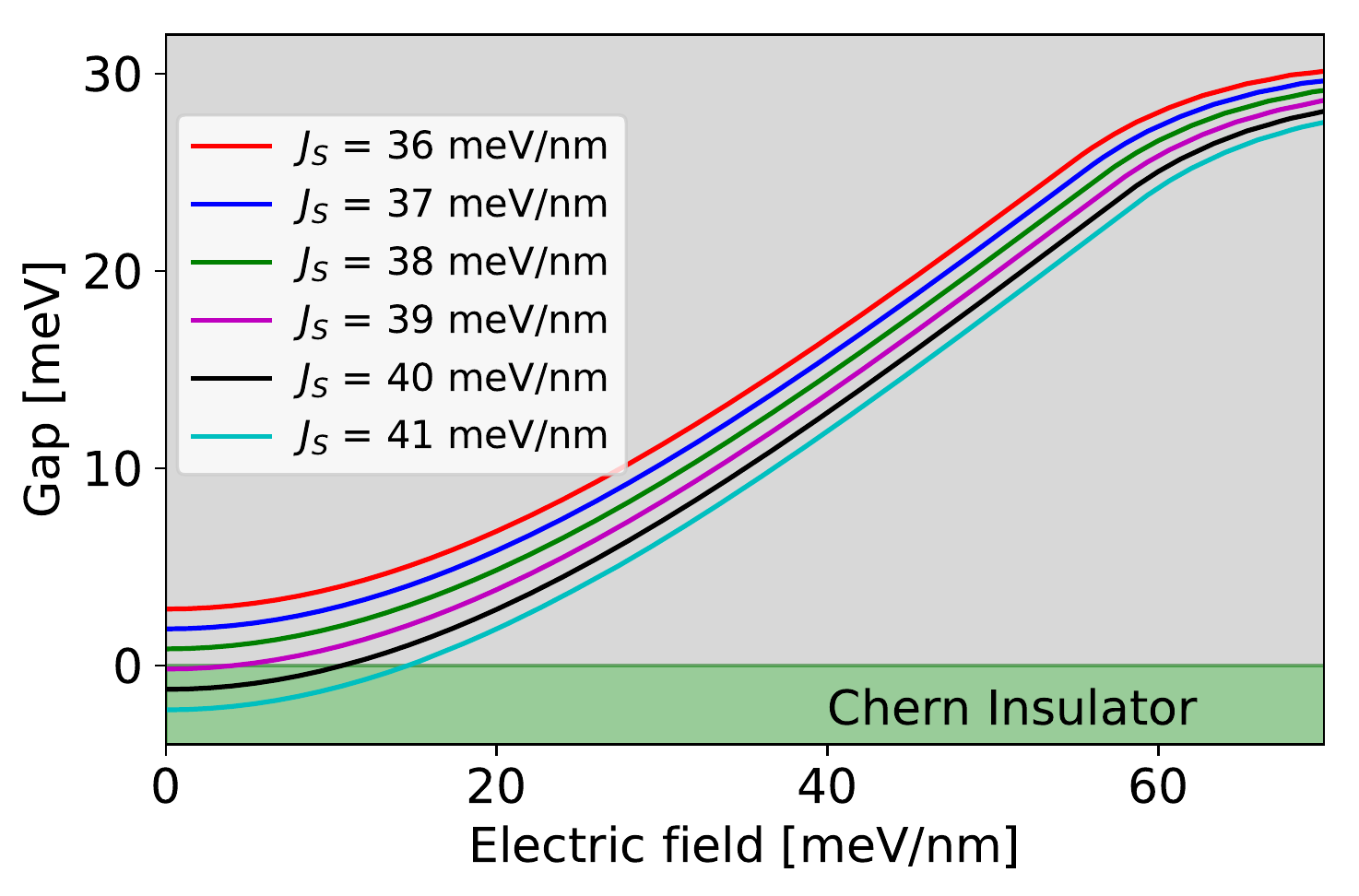}
  \caption{Gap {\it vs.} electric field of trilayer(N=3) MnBi$_2$Te$_4$ thin film with different value of exchange interaction $J_S$, which is varied from 36 meV to 41 meV. The topoloigcal phase transition happens at $J_S \approx 39~ meV$ in the absence of electric field, while in the range of Chern insulator phase, a small electric field would drive the thin film into a trivial insulator phase.
  }\label{fig:gap_trilayer}
\end{figure}
\fi

\section{Charge distribution}
Fig. \ref{fig:hartree_charge} are the charge distribution at the neutrality points for antiferromagnetic MBT thin films, with the total charge density to be 0 and the two surfaces of thin films are electrically polarized. 

\ifpdf
\begin{figure}[htp]
  \centering
  \includegraphics[width=0.9 \linewidth ]{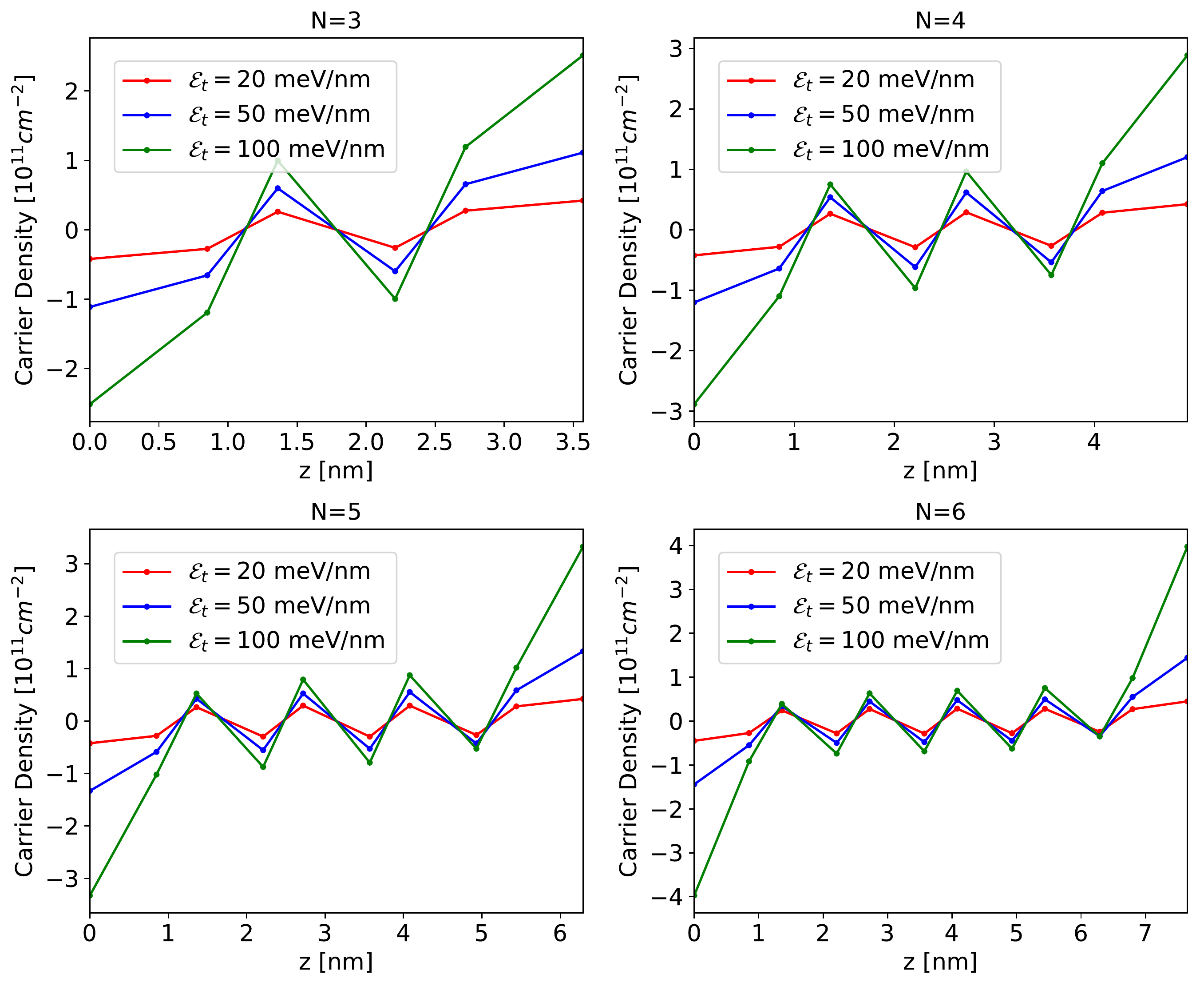}
  \caption{Charge distribution at the neutrality points for antiferromagnetic MBT thin films, (a)-(d) are corresponding to the case of N = 3-6. 
  }\label{fig:charge}
\end{figure}
\fi

The electric field distributions across the antiferromagnetic MBT thin films are shown in Fig. \ref{fig:efield}, (a)-(d) are showing the cases with thickness from 3 to 6 septuples.

\ifpdf
\begin{figure}[htp]
  \centering
  \includegraphics[width=0.9 \linewidth ]{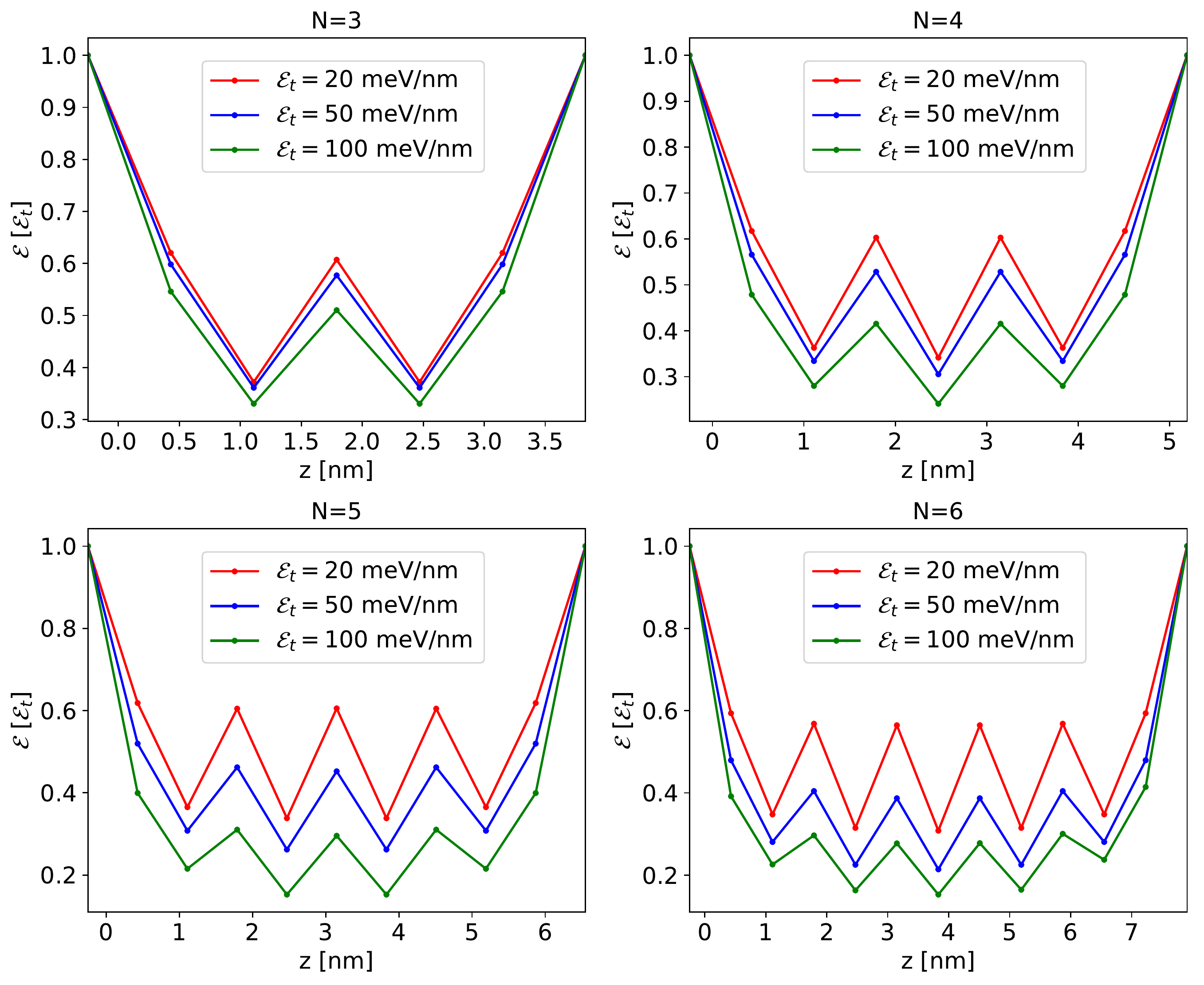}
  \caption{Electric field distribution of acrossing the antiferromagnetic MBT thin films with different external field. In these plots the electric fields are ploted with the unit of the external electric field $\mathcal{E}_t$, the field inside the thin films decrease due to the screening.
  }\label{fig:efield}
\end{figure}
\fi

Due to the screening, the electric fields inside the thin films decrease. In Fig. \ref{fig:dielectric_constant} we estimate the effective dielectric constants with the methods as follow:
\begin{equation}
    \epsilon_{eff} = \frac{\mathcal{E}_t \times d \times N}{V_{b} - V_t},
\end{equation}
where ${E}_t$ is the eleectric field at the top surface, $V_{t/b}$ is the Hartree potential at the top/bottom surface, and d is the thickness of the thin films.

\ifpdf
\begin{figure}[htp]
  \centering
  \includegraphics[width=0.9 \linewidth ]{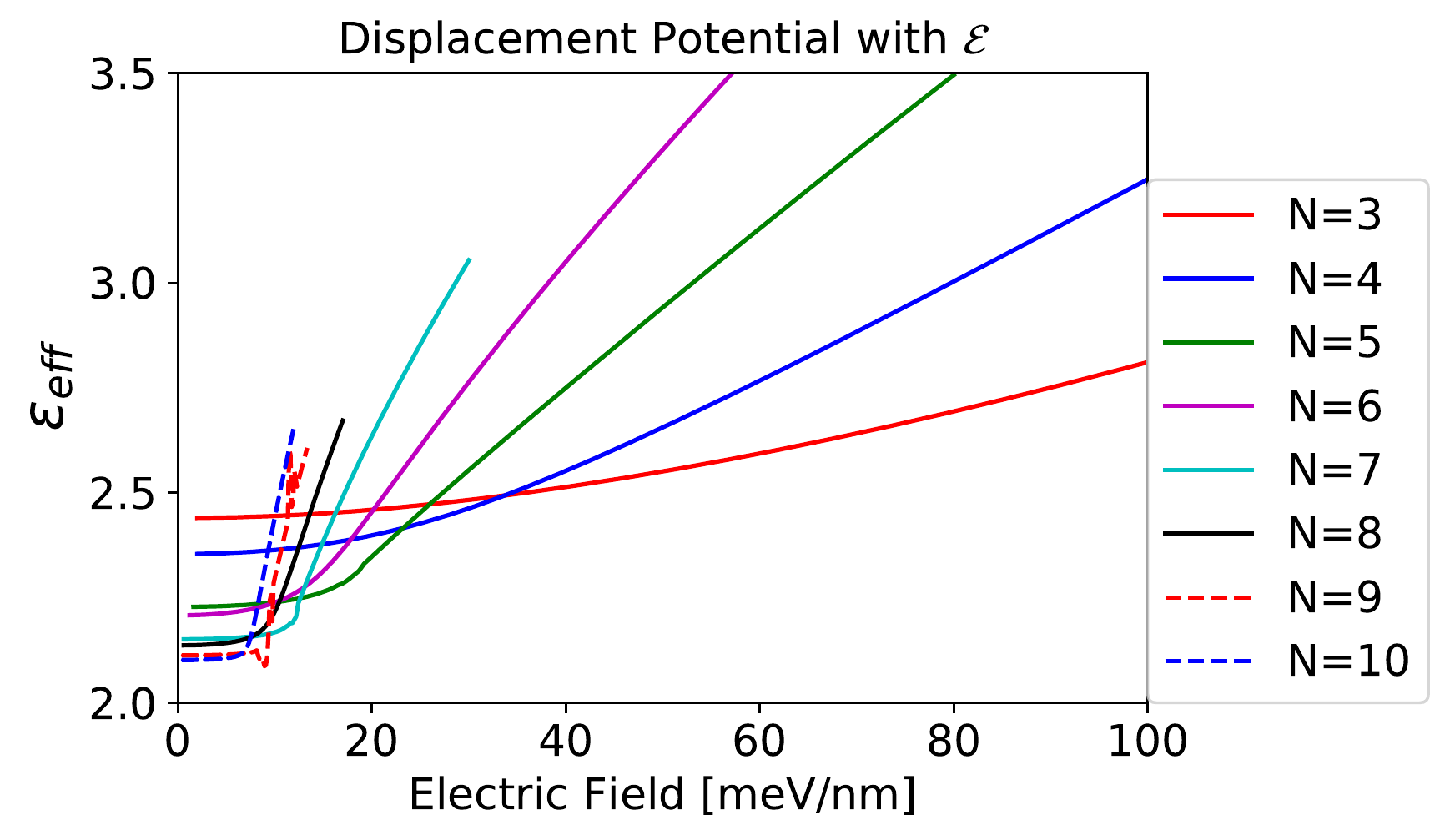}
  \caption{Effective dielectric constant {\it vs.} electric fields for MBT thin film with thickness of N = 3-10.
  }\label{fig:dielectric_constant}
\end{figure}
\fi





\section{Gap {\it vs.} Electric Field }

The gaps at  $\Gamma$ points from both DFT calculations and the couple Dirac cone model are shown in Fig. \ref{fig:gap_field_dft_model}, (a) and (b) are the case from DFT calculations while (c) and (d) are got from the couple Dirac cone model. In these plots we assign a minus sign for the thin films with Chern number of 1.
\ifpdf
\begin{figure}[htp!]
  \centering
  \includegraphics[width=0.9 \linewidth ]{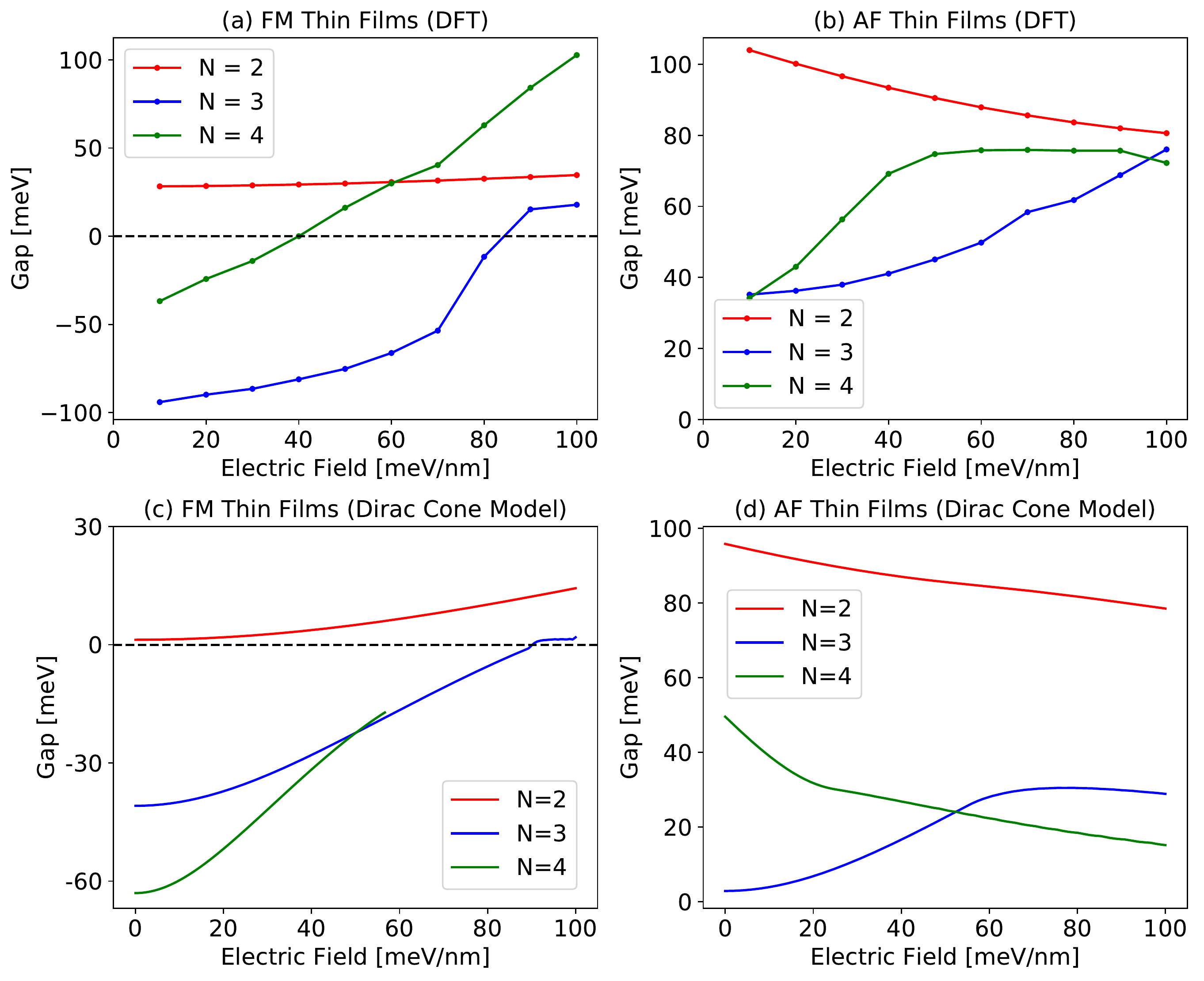}
  \caption{
  Dependence of gap of thin films {\it vs.} external electric field, (a) and (b) are the case from DFT calculations while (c) and (d) are got from the couple Dirac cone model.
  }\label{fig:gap_field_dft_model}
\end{figure}
\fi

\section{Toy model for even-odd effect in AF thin films}
The antiferromagnetic MBT thin films may be modeled by a toy model, thin films with odd and even number of layers may be modeled by a topological insulator thin film with the same mass and opposite mass term within the Dirac cone at the two surface. 

For MBT thin films with odd number of layers, the two surface Dirac cones have the same mass term, the effective Hamiltonian in the absence of electric fields is:
\begin{equation}
    H_{odd} = \begin{pmatrix}
m & v_D k^{+} & \Delta & 0\\
v_D k^{-} & -m & 0 & \Delta \\
\Delta  & 0 & m & -v_D k^{+}\\
0 & \Delta  & -v_D k^{-} & -m\\
\end{pmatrix},
\end{equation}
where $k^{\pm} = k_x \pm i k_y$, $v_D$ is the Dirac velocity, $\Delta$ is the hybridization of the two Dirac-cone at the surface and $m$ is the mass. 
$\Delta$ is the effective hybridization which may be got from the parameters $\Delta_S$ and $\Delta_D$ in the main text, and $m$ may be approximately only dependent on $J_S$.
The eigenvalues are:
\begin{equation}
    E = \pm \sqrt{v_D^2 |k|^2 + (m \pm \Delta)^2}, 
\end{equation}
where $|k| = \sqrt{k_x^2 + k_y^2}$.
In the presence of electric field, a perturbation Hamiltonian may be modeled as:
\begin{equation}
V_{H} = \begin{pmatrix}
V & 0 & 0 & 0\\
0 & V & 0 & 0 \\
0  & 0 & -V & 0\\
0 & 0 & 0 & -V\\
\end{pmatrix},
\end{equation}
with $\pm V$ the Hartree potential due to the presence of electric field.

For MBT with even number of layers, the effective Hamiltonian in the absence of electric field is:
\begin{equation}
    H_{even} = \begin{pmatrix}
m & v_D k^{+} & \Delta & 0\\
v_D k^{-} & -m & 0 & \Delta \\
\Delta  & 0 & -m & -v_D k^{+}\\
0 & \Delta  & -v_D k^{-} & m\\
\end{pmatrix},
\end{equation}
in which the two Dirac cones at the two surfaces have oppisite mass.
The eigenvalues are:
\begin{equation}
    E = \pm \sqrt{v_D^2 |k|^2 + m^2 + \Delta^2},
\end{equation}



\section{Phase transition {\it vs.} electric field}
The topological phase transition point may be got via the band energy at $\Gamma$ point, for MBT thin films with odd number of layers, the eigenvalues at $\Gamma$ point are:
\begin{equation}
    E = \pm m \pm \sqrt{\Delta^2 + V^2},
\end{equation}
where V is the Hartree potential caused by the electric field. In the absent of electric field, the thin films are in topological phase when $m>\Delta$. In the presence of electric field, this condition becomes as $m> \sqrt{\Delta^2 + V^2}$, which means the exchange interaction needed to induce a topological phase transition is larger and increase with the increase of electric field.
While for MBT thin films with even number of layers, the eigenvalues at $\Gamma$ point are:
\begin{equation}
    E = \pm \sqrt{(m \pm V)^2 + \Delta^2},
\end{equation}
in this case the there is still no topological phase transition, similar with the case in the absent of electric field.


\end{document}